\documentclass{aa}  
\usepackage{graphicx}
\usepackage{txfonts}
\begin{document}
   \title{Quasi-classical rate coefficient calculations for the
    rotational (de)excitation of H$_2$O by H$_2$}

   \author{A. Faure
     \inst{1}
     \and
     N. Crimier\inst{1}
     \and
     C. Ceccarelli\inst{1}
     \and
     P. Valiron\inst{1}
     \and
     L. Wiesenfeld\inst{1}
     \and
     M.~L. Dubernet\inst{2}
   }

   \offprints{A. Faure}

     \institute{Laboratoire d'Astrophysique, UMR 5571 CNRS, Universit\'e
     Joseph-Fourier, B.P. 53, 38041 Grenoble cedex 09, France \\
     \email{afaure@obs.ujf-grenoble.fr}
   \and
     Observatoire de Paris-Meudon, LERMA UMR 8112 CNRS, 5 Place Jules Janssen, 
     92195 Meudon Cedex, France \\
       }
   \date{Received 20 April 2007 / Accepted 4 June 2007}

 
  \abstract
      {The interpretation of water line emission from existing observations 
	and future HIFI/Herschel data requires a detailed knowledge of 
	collisional rate coefficients. Among all relevant
	collisional mechanisms, the rotational (de)excitation of H$_2$O
	by H$_2$ molecules is the process of most interest in 
	interstellar space.}
      {To determine rate coefficients for rotational de-excitation among the 
	lowest 45 para and 45 ortho rotational levels of H$_2$O colliding
	with both para and ortho-H$_2$ in the temperature range 20$-$2000~K.}
      {Rate coefficients are calculated on a recent high-accuracy
	H$_2$O$-$H$_2$ potential energy surface using quasi-classical
	trajectory calculations. Trajectories are sampled by a canonical Monte-Carlo 
	procedure. H$_2$ molecules are assumed to be rotationally thermalized 
	at the kinetic temperature.}
      {By comparison with quantum calculations available for low lying levels, 
	classical rates 
	are found to be accurate within a factor of 1$-$3 for the dominant transitions, 
	that is those with rates larger than a few 10$^{-12}$cm$^{3}$s$^{-1}$. 
	Large velocity gradient modelling shows that the new rates have a significant 
	impact on emission line fluxes and that they should be adopted in any detailed 
	population model of water in warm and hot environments.}
      {}

   \maketitle
%

\section{Introduction}

Since its discovery in interstellar space (Cheung et
al. \cite{cheung69}), water vapour has been detected in a great
variety of astronomical objects using both Earth-based and, foremost,
spacecraft observations. The {\it Infrared Space Observatory} (ISO),
in particular, has revealed the ubiquity of water in the interstellar
(ISM) and circumstellar (CSM) media (for a recent review see
Cernicharo \& Crovisier \cite{cernicharo05}). The more recent {\it
Submillimetre Wave Astronomy Satellite} (SWAS) and {\it Odin} missions
have also measured gaseous water in a wide variety of sources using
the single ground-state rotational transitions of ortho-H$_2$O at
557~GHz (e.g. Melnick \& Bergin \cite{melnick05}; Hjalmarson et
al. \cite{hjalmarson03}). Finally, water is also detected at radio and
(sub)millimeter wavelengths through maser transitions which are
commonly associated with star-forming regions (e.g. Cernicharo et
al. \cite{cernicharo90,cernicharo99}; Goddi et al. \cite{goddi07} and
references therein or extraglactic sources (e.g. Cernicharo et
al. \cite{cernicharo06}; Kondratko et al. \cite{kondratko06} and
references therein).

The abundance of water varies largely in cold and warm media. In the
former, because of the freezing of water onto the dust grains, water
does not exceed $\sim10^{-8}$ (Bergin \& Snell \cite{bergin02};
Poelman et al. \cite{poelman07} and references therein). But in warm
regions, water can become the most abundant molecule after H$_2$, with
an abundance of $\sim10^{-4}$. In star forming regions, this occurs
both because of the release in the gas phase of the grain ices and
because of endothermic reactions efficiently forming water in the gas
phase (e.g. Ceccarelli et al. \cite{ceccarelli96}; Doty \& Neufeld
\cite{doty97} and references therein). Under these conditions, water
becomes a crucial molecule in the thermal and chemical balance.  In
fact, owing to its large dipole, water lines will dominate the cooling
(and sometime even heating) of the gas in a large range of gas
densities and temperatures (e.g. Neufeld \& Melnick \cite{neufeld91};
Ceccarelli et al. \cite{ceccarelli96}; Cernicharo et
al. \cite{cernicharo06} and references therein).  Also, since oxygen
will be mostly locked in water molecules at temperatures larger than
about 200~K, the chemical abundance of more complex and less abundant
molecules will depend on the water abundance.

For all these reasons, water is a key molecule in space, and its
observation and the determination of its abundance are critical for
many theoretical aspects.  Indeed, observing and measuring the water
lines and their abundance is an important task for the high resolution
spectrometer {\it Heterodyne Instrument for the Far Infrared} (HIFI)
on board the ESA funded Herschel satellite, which will be launched in
2008. However, as stated by Cernicharo \& Crovisier
(\cite{cernicharo05}), ``little new information will be obtained if
collisional rates adapted to the temperatures of the clouds in the ISM
and CSM are not available''.

Collisional rate coefficients are indeed essential in the description
of the energy exchange processes responsible for molecular line
formation in astronomical environments. Spectral features such as
maser emission are produced in low-density conditions far from
thermodynamic equilibrium and through a complex competition between
radiative and collisional processes, including infrared emission from
dust (e.g. Poelman et al. \cite{poelman07}). In the ISM and CSM, the
main colliding partners are hydrogen molecules, hydrogen and helium
atoms, and electrons. In general, however, H$_2$ is the dominant form
of hydrogen and as it is more abundant than He (by a factor of $\sim
5$) and electrons (by $\sim$ 4$-$8 orders of magnitude), H$_2$ is
usually the dominant exciting species. At typical ISM and CSM
temperatures ($T<2000$~K), the dominant energy exchange process is
thus rotational (de)excitation by H$_2$, although vibrational
excitation cannot be neglected in the highest temperature and density
regions.

To the best of our knowledge no experimental state-to-state data is
available for H$_2$O$-$H$_2$ except for the total relaxation rate of
the water bending mode (see Faure et al. \cite{faure05a} and
references therein). On the theoretical side, quantum close-coupling
(CC) state-to-state calculations for the rotational excitation of
H$_2$O by H$_2$ and He have been performed for kinetic temperatures
and energy levels lower than 140~K and 2000~K, respectively (Green et
al. \cite{green93}; Phillips et al. \cite{phillips96}; Dubernet et
al. \cite{dubernet02}; Grosjean et al. \cite{grosjean03}; Dubernet et
al. \cite{dubernet06}). Hence, current models of water emission in
warm and hot environments rely exclusively on rates for excitation by
He atoms, usually scaled by the reduced mass ratio $\rm
(\mu_{H_2O-H_2}/\mu_{H_2O-He})^{1/2}$ (e.g. Poelman \& Spaans
\cite{poelman05}). It has been shown by Phillips et
al. (\cite{phillips96}), however, that excitation by para-H$_2$ in its
$J$=0 level is not too different from excitation by He atoms (with
most rates differing by a multiplicative factor of 1$-$3) but that
excitation by H$_2$ in excited rotational levels is significantly
different. This mainly reflects the importance of the long-range
interaction between the H$_2$ quadrupole moment and the dipole of
H$_2$O (Phillips et al. \cite{phillips96}). Rates for excitation of
H$_2$O by H$_2$ at high temperatures and/or for high rotational levels
are therefore urgently needed for the interpretation of existing and
future water emission spectra.

In principle, scattering calculations based on the quantum CC method
can provide an absolute accuracy of a few percent for a given
potential energy surface (PES). As an illustrative example, the recent
study of Gilijamse et al. (\cite{gilijamse06}) has shown an excellent
agreement between experimentally measured cross sections and CC
calculations based on {\it ab initio} PES. The major drawback of the
quantum CC method is its computational cost which increases
dramatically with the number of coupled channels, i.e. with the total
energy and the number of degrees of freedom. Very recently, Dubernet
et al. (\cite{dubernet06}) provided new rotational rate coefficients
for H$_2$O$-$H$_2$ at temperatures below 20~K using fully converged CC
calculations based on the recent high-accuracy PES of Faure et
al. (\cite{faure05a}). The main objective of their study was to test
the influence of the new PES on the scattering calculations and to
make comparisons with previous results based on the PES of Phillips et
al. (\cite{phillips94}). A significant re-evaluation of rates was
observed, especially for para-H$_2$, with differences up to a factor
of 3 at 20~K. Full CC calculations, augmented with coupled-states and
infinite order sudden (IOS) calculations, are currently in progress at
higher collision energies with the objective of obtaining quantum CC
rates for kinetic temperatures and rotational levels up to
$\sim$~2000~K and making comparisons with pressure broadening
experiments (Dubernet et al., in preparation).

The present study is motivated by the possibility of performing high
temperature quasiclassical trajectory (QCT) calculations as an
alternative to quantum mechanical calculations. In contrast to CC
calculations, the computational time required for QCT calculations
decreases as the collisional energy increases and such calculations
are thus particularly adapted to high temperatures where purely
quantum effects are usually negligible. QCT calculations have been
shown to accurately reproduce quantum results provided that the
relevant transition probabilities are large enough. In particular,
Faure et al. (\cite{faure06}) recently observed a good agreement
between classical and quantum rates for the rotational de-excitation
of H$_2$O by H$_2$ at a single temperature of 100~K. The average
accuracy of classical results was found to be better than a factor of
2 for the transitions with the largest probabilities, i.e. those with
a rate greater than $\sim$ 10$^{-11}$cm$^{3}$s$^{-1}$. The classical
approach is thus expected to provide a good compromise between
accuracy and computational effort at temperatures where converged
quantum CC calculations are extremely expensive (typically
$T>$300~K). The next section describes details of the QCT
calculations. Rate coefficients are presented in Sect.~3. A first
application of these rates is given in Sect.~4. Conclusions are drawn
in Sect.~5.


\section{Quasi-classical trajectory method}

\subsection{Potential energy surface and general procedure}

All calculations presented below were performed with rigid molecules
using the vibrationally averaged PES of Faure et
al. (\cite{faure05a}), where full details can be found. The angular
expansion of the PES was obtained as a least square fit of thousands
of {\it ab initio} values over 149 angular functions which are
specifically adapted to quantum calculations and provide analytical
derivatives for the classical equations of motion. In order to reduce
computational time in integrating these equations, a subset of only 45
angular functions was retained. This selection was found to reproduce
the {\it ab initio} values within a few cm~$^{-1}$ in the Van der
Waals minimum region. The inclusion of additional angular functions
was tested through trajectory analysis and found to be negligible. The
water and hydrogen geometries were taken at their effective rotational
values, that is those corresponding to the spectroscopically
determined rotational constants.

As inelastic (nonreactive) collisions cannot interconvert the ortho-
and para-forms of either species, QCT calculations were done
separately for the four nuclear spin combinations, namely
para-H$_2$O$-$para-H$_2$, para-H$_2$O$-$ortho-H$_2$,
ortho-H$_2$O$-$para-H$_2$ and ortho-H$_2$O$-$ortho-H$_2$. Classical
trajectories were run using the Monte-Carlo procedure described in
Faure et al. (\cite{faure06}). The rate coefficients were computed by
a canonical sampling of the initial Maxwell-Boltzmann distribution of
collisional energy at four kinetic temperatures, namely $T=$100, 300,
1000 and 2000~K. Some test calculations were also performed at 50~K
(see Fig.~1). The classical equations of motion were numerically
solved using an extrapolation Bulirsch-Stoer algorithm. The total
energy was conserved up to six digits, i.e. within
$\sim$0.1~cm$^{-1}$. The maximum impact parameter $b_{\rm max}$,
defined as the smallest impact parameter for which 1000 trajectories
produced no change in the rotational state of H$_2$O, was found to
range between 10 and 16~a.u. at the selected temperatures. Batches of
10,000 trajectories were run for each of the lowest 44 para and 44
ortho excited rotational levels of H$_2$O, resulting in a total of
7,040,000 trajectories. These calculations required approximately
150~hours of computer time on {\it AMD Opteron} quadri-processors for
each temperature and for each of the four nuclear spin cases.

\subsection{Initial state selection}

For each molecule, the classical rotational angular momentum was taken
as $[J(J+1)]^{1/2}$, where $J$ is the rotational quantum (integer)
number, using the correspondence principle. For water, initial values
of the classical angular momentum components were simply assigned to
the pseudoquantum numbers $K_a$ and $K_c$ (projection of $J$ along the
axis of the least and greatest moments of inertia, respectively). This
trivial initial selection was found to reproduce exact quantum
eigenvalues within a few cm$^{-1}$ for the lowest levels (see Table~1
in Faure et al. \cite{faure06}) and within a few percent for the
highest. For H$_2$, the initial para and ortho rotational levels were
weighted separately according to the Boltzmann distribution:
\begin{equation}
  \rho(J_2)=\frac{(2J_2+1)\exp(-\frac{E_{J_2}}{kT})}{\sum_{J_2}
  (2J_2+1)\exp(-\frac{E_{J_2}}{kT})},
\end{equation}
where $J_2$ is the H$_2$ angular momentum, $k$ is the Boltzmann
constant and the sum in the denominator extends only over even (odd)
numbers for para (ortho) H$_2$. We thus assumed that each nuclear spin
species of H$_2$ has a thermal distribution of rotational levels. This
is obviously not necessarily the case in astronomical environments, in
particular for high-$J_2$ levels. However, from the H$_2$ Einstein
coefficients tabulated by Wolniewicz et al. (\cite{wolniewicz98}) and
the H$_2$-He rates reported by Flower et al. (\cite{flower98}), we
have estimated the He critical densities to be below $\sim$
10$^3$~cm$^{-3}$ up to $J_2=5$, that is up to H$_2$ levels lying more
than $\sim$2000~K above the ground states. These levels are therefore
likely to be at or close to local thermodynamic equilibrium (LTE) at
typical ISM and CSM densities. Observationally, the detections of
H$_2$ in shocks by Lefloch et al. (\cite{lefloch03}) and Neufeld et
al. (\cite{neufeld06}) also suggest that the H$_2$ rotational
populations are close to LTE for $J_2$ as high as 9. As the dependence
of the H$_2$O rates on the H$_2$ rotational level is modest (see
Sect.~3), the assumption of rotationally thermalized H$_2$ molecules
seems reasonable. Furthermore, this assumption has the advantage that
the rate coefficients rigorously satisfy the principle of detailed
balance (see below).

\subsection{Final state selection and rate calculation}

In the case of an asymmetric top molecule, the standard bin histogram
method used to assign the final classical action has been shown to be
ambiguous (Faure \& Wiesenfeld \cite{faure04b}). However, in the
particular case of water-like molecules, i.e. asymmetric rotors with
{\it i)} dipole along the axis of intermediate moment of inertia and
{\it ii)} $C_{\rm 2V}$ symmetry, the ambiguities can be solved by
taking advantage of the fact that ortho/para transitions are not
permitted (see Faure et al. \cite{faure06}). For H$_2$, no final
assignment was required as rate coefficients must be summed over all
possible H$_2$ rotational levels. The rate coefficients reported in
the present study are therefore defined, for each of the four spin
combinations, as:
\begin{equation}
  R(J_{\alpha} \rightarrow J'_{\alpha'}) = \sum_{J_2,
  J'_2}\rho(J_2)\,R(J_{\alpha}, J_2 \rightarrow J'_{\alpha'}, J'_2),
\end{equation}
where $\alpha=K_a, K_c$. According to the canonical Monte-Carlo
procedure described above, these rates were computed as:
\begin{equation}
   R(J_{\alpha} \rightarrow J'_{\alpha'}, T) =
  \left(\frac{8kT}{\pi\mu}\right)^{1/2}\pi b_{\rm
  max}^2\frac{N}{N_{\rm tot}},
\end{equation}
where $\mu$ is the reduced mass, $N$ is the number of trajectories
leading to a specified final level of water $J'_{\alpha'}$ and $N_{\rm
tot}$ is the total number of 'physical' trajectories\footnote{A small
fraction (less than 10\%) of trajectories are indeed rejected due to
either non-conservation of energy (failure of numerical integration)
or unphysical final values of $K_a, K_c$.}, typically between 9,000
and 10,000. The Monte-Carlo standard deviation is:
\begin{equation}
\frac{\Delta R(J_{\alpha} \rightarrow J'_{\alpha'})}{R(J_{\alpha}
\rightarrow J'_{\alpha'})}= \left(\frac{N_{\rm tot}-N}{N_{\rm
tot}N}\right)^{1/2}.
\end{equation}
In the following, statistical errors are defined as two standard
deviations which, in practice, correspond to a typical statistical
accuracy of 10$-$30\%.

\subsection{Detailed balance}

The principle of detailed balance (microsopic reversibility), which is
the consequence of the invariance of the interaction under time
reversal, states that the rate coefficients, as defined in Eq.~(2),
must obey the relation:
\begin{equation}
  R(J'_{\alpha'} \rightarrow J_{\alpha}, T)=\frac{(2J+1)}{(2J'+1)}\exp
  \left[\frac{E_{J'_{\alpha'}}-E_{J_{\alpha}}}{kT}\right]R(J_{\alpha}\rightarrow
  J'_{\alpha'}, T).
\end{equation}
As shown by Faure et al. (\cite{faure06}), however, QCT calculations
significantly overestimate the excitation (endoergic) rates in the
case of H$_2$O transitions whose energies are larger or comparable to
the kinetic temperature. This `threshold effect' results from the
rotational degrees of freedom of H$_2$O that are classically active at
energies below quantum thresholds. The competing effect induced by the
rotational degrees of freedom of H$_2$ was also observed, to a lesser
extent, in the case of H$_2$O transitions whose energies are lower
than the kinetic temperature. Such effects were also reported for
other hydrogenic systems (see Mandy \& Pogrebnya \cite{mandy04} and
references therein). As a result, detailed balance is not rigorously
satisfied at the classical level and excitation rates can be
ovestimated by an order of magnitude. The recommended practical
procedure is therefore to employ QCT calculations for de-excitation
(exoergic) transitions and to apply Eq.~(5) to obtain the reverse,
endoergic, rates. We note, however, that detailed balance was actually
satisfied within error bars for a number of low energy transitions at
the highest temperatures\footnote{It should be noted that `effective'
rates which depend on the initial rotational levels of H$_2$ and which
are summed over the final H$_2$ rotational levels (as defined in
Eq.~(1) of Dubernet et al. \cite{dubernet06}) do not obey detailed
balance since they do not involve a complete thermal average. These
effective rates were reported in the previous quantum studies.}.

\subsection{Advantages, shortcomings and accuracy of the classical method}

The shortcomings of the classical approach are well known and are
usually marked at low kinetic temperatures where purely quantum
effects are significant. In particular, between $\sim$ 1 and 100~K
where the Van der Waals interactions are dominant, quantum rates
exhibit oscillatory temperature dependence due to the presence of
quantum resonances (of shape and/or Feshbach type). In this regime,
classical calculations are obviously questionable. Amplitudes of these
oscillations are however usually modest and not much larger than the
accuracy of QCT calculations. We note, for instance, that the recent
study of Wernli et al. (\cite{wernli07}) showed a quite good agreement
between QCT and quantum rates down to 10~K. A greater limitation of
the classical approach is the quantization of the rotational degrees
of freedom, as explained above. Finally, a major disadvantage of
classical mechanics is its inability to reproduce small probabilities,
the so-called classically forbidden transitions corresponding to rates
lower than $\sim$10$^{-13}- 10^{-12}$cm$^{3}$s$^{-1}$. Owing to all
these intrinsic limitations that add to the Monte-Carlo statistical
deviations, an accuracy of greater than 10\% is difficult to achieve
by QCT rate calculations. The recent work of Faure et
al. (\cite{faure06}) has shown, however, that classical rates for
H$_2$O$-$H$_2$ are accurate within a multiplicative factor of 1$-$3
for low lying levels. Precision of $\sim$30\% was even observed for
the greatest rates. Such accuracies are still much lower than the few
percent precision of quantum CC calculations.

However, a major advantage of classical trajectories is to allow the
computation of de-excitation rates for high initial values of the
rotational angular momenta. In particular, the calculation of rate
coefficients as defined in Eq.~(2) requires the inclusion of all
significantly populated levels of H$_2$. At 2000~K and assuming LTE,
the most populated level is $J_2$=4 (3) for para(ortho) H$_2$ but
levels up to $J_2$=14 (13) were in fact included in our QCT
calculations. At the quantum CC level, a calculation with $J_2>4$ is
currently not feasible. Thus, classical mechanics proves particularly
suited to collisions between `hot' molecules where standard quantum
methods are not applicable.

\section{Quasi-classical trajectory results}

\subsection{Comparison to quantum data}

In Fig.~1, four sets of rate coefficients for rotational de-excitation
of the ortho-H$_2$O $3_{03}$ level by para and ortho-H$_2$ are plotted
as a function of temperature: {\it i)} current QCT data between 50 and
2000~K, {\it ii)} quantum data of Phillips et al. (\cite{phillips96})
between 20 and 140~K, {\it iii)} H$_2$O$-$He quantum data of Green et
al. (\cite{green93}) multiplied by the reduced mass ratio $\rm
(\mu_{H_2O-H_2}/\mu_{H_20-He})^{1/2}=1.348$ between 20 and 2000~K,
{\it iv)} quantum data of Dubernet et al. (\cite{dubernet06}) at 20~K
augmented with unpublished data for para-H$_2$ between 20 and
1500~K. These latter unpublished rates have been computed using
Eq.~(2) with the sum extending over $J_2=0, 2, 4$ and with a scaling
procedure for estimating the rates for excitation by H$_2(J_2=4)$ from
those by H$_2(J_2=2)$\footnote{A scaling factor of 1.2 has been
uniformly employed for all transitions.}. We have selected the
$3_{03}$ level because it is the highest initial state included by
Phillips et al. (\cite{phillips96}). Firstly, it can be observed that
there is an overall good agreement (within a factor of 1$-$3) between
the present rates and those of both Phillips et
al. (\cite{phillips96}) and Dubernet et al. while large differences
(up to 2 orders of magnitude) are found with the scaled H$_2$O$-$He
rates, especially so for ortho-H$_2$ at $T<$100~K. It should be noted
in particular that at low temperatures, the transition $3_{03}-1_{10}$
has a larger rate than $3_{03}-2_{21}$ for para-H$_2$ while the
reverse holds for ortho-H$_2$. Second, in the case of para-H$_2$, the
agreement between the present classical calculations and the quantum
calculations of Dubernet et al. is excellent above 100~K, where
quantum data are within the error bars of the QCT data. At and below
100~K, the agreement is not as good, reflecting the limitations of the
classical approach (see Sect.~2.5), but differences do not exceed a
factor of 2. Third, the influence of the new H$_2$O$-$H$_2$ PES is
modest, except for para-H$_2$ at low temperature. Classical rates for
ortho-H$_2$ are thus found to be in very good agreement with those of
Phillips et al. (\cite{phillips96}). This small impact of the PES on
the ortho-H$_2$ rates was previously observed by Dubernet et
al. (\cite{dubernet06}). Fourth, we observe that the differences
between the present classical rates and the scaled H$_2$O$-$He rates
decrease with increasing temperature. This reflects the fact that at
high collision energy, the scattering process becomes dominated by
kinematics rather than specific features of the PES. Finally, above
300~K, the difference between para and ortho-H$_2$ rates becomes
minor. This clearly shows that the dependence of the rates on the
H$_2$ rotational level is weak, provided that $J_2\geq 1$. We have
actually checked that differences between de-excitation rates of the
$3_{03}$ level by H$_2(J_2=1)$ and H$_2(J_2=6)$ do not differ by more
than four Monte-Carlo standard deviations, i.e. by less than a factor
of 2. This finding is consistent with the quantum results of Phillips
et al. (\cite{phillips96}) who found similar rates for excitation of
para-H$_2$O by H$_2(J_2=1)$ and H$_2(J_2=2)$.

\begin{figure}[h]
  \centering
  \resizebox{\hsize}{!}{\includegraphics*[angle=-90]{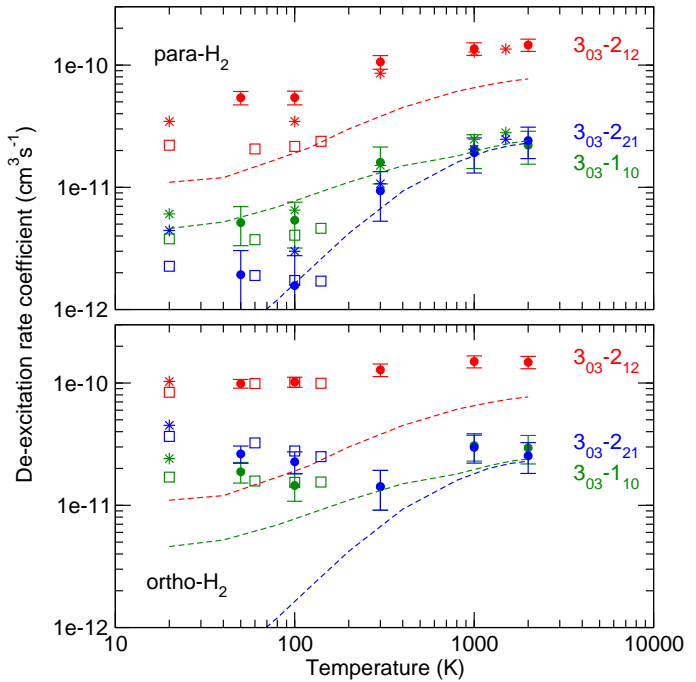}}
  \caption{Rate coefficients, Eq.~(2), for rotational de-excitation of
  the ortho-H$_2$O state $3_{03}$ as functions of temperature for
  para-H$_2$ (upper panel) and ortho-H$_2$ (lower panel). Squares
  denote quantum results by Phillips et al. (\cite{phillips96}); stars
  refer to the quantum calculations of Dubernet et al.; dashed lines
  denote quantum (scaled) results of Green et al. (\cite{green93}) for
  H$_2$O$-$He; circles with error bars (95\% statistical confidence
  level) give the present QCT results. See text for details. Note that
  the transition $3_{03}\rightarrow 1_{01}$ is not plotted for
  clarity.}
  \label{Figrates1}
\end{figure}

\subsection{Accuracy and propensity rules}

The above observations obviously hold for other initial levels and,
from numerous comparisons, we estimate the QCT results to be accurate
within a factor of 1$-$3 for the largest rates, i.e. those larger than
a few 10$^{-12}$cm$^{3}$s$^{-1}$, as observed in Faure et
al. (\cite{faure06}). For smallest rates, errors larger than a factor
of 3 can be encountered, illustrating the inability of classical
mechanics to reproduce small probability events. As a result, rates
less than $\sim$ 10$^{-12}$cm$^{3}$s$^{-1}$, which correspond to very
poor statistics (zero or less than $\sim$10 events over 10,000), were
set to zero. This means that a significant number of transitions, the
so-called classically forbidden or {\it rare} transitions, have null
classical rates. Moreover, some transitions were found to have null
rates at 100~K but significant rates at higher temperatures. Our
strategy to tackle this problem was to retain transitions obeying the
following two criteria: the rate must be {\it i)} larger than
10$^{-12}$cm$^{3}$s$^{-1}$ at 300~K and above and {\it ii)} larger
than 10$^{-11}$cm$^{3}$s$^{-1}$ at 2000~K. If one of the two criteria
is not fulfilled, then the transition is classified as a rare event
and the corresponding (scaled) H$_2$O$-$He rate is adopted. It should
be noted that these rare events correspond in most cases to
transitions with $\Delta J>3$ which are not expected to play a
dominant role in the radiative transfer equations (see
Sect.~4). Moreover, despite the fact that no rigorous selection rules
hold for molecule-molecule collisions, we actually observed the
following {\it propensity} rules:
\begin{equation}
\Delta J=0, \pm 1; \hspace{0.5cm} \Delta K_a=0, \pm 1; 
\hspace{0.5cm} \Delta K_c=0, \pm 1,
\end{equation}
which are consistent with the classical picture of a minimum
reorientation of the angular momentum of a strongly asymmetric-top
molecule\footnote{In the case of a slightly asymmetric-top such as
H$_2$CO, $K_a$ and $K_c$ are further constrained by the value of
$J$.}. These rules are illustrated in Fig.~1 where the transition
$3_{03}-2_{12}$ is indeed found to be strongly favoured.

\subsection{Fitting procedure}

For use in astrophysical modelling, the transition rates
$R(J'_{\alpha'} \rightarrow J_{\alpha}, T)$ must be evaluated on a
sufficiently fine temperature grid. For this, the rates were first
least-square fitted over 100$-$2000~K by the analytic form used by
Faure et al. (\cite{faure04}):
\begin{equation}
\log_{10}R(J'_{\alpha'} \rightarrow J_{\alpha}, T)=\sum_{n=0}^N
a_nx^n,
\end{equation}
where $x=T^{-1/6}$. A typical fitting accuracy of a few percent was
obtained. In the case of rare events with large error bars, the
fitting method was also found to smooth the temperature dependence of
the rates and thus remove irregularities caused by poor
statistics. Note that $T=$50~K was investigated only for a few levels
and was not retained in the fitting procedure. We then evaluated the
fitted rates at the following grid temperature values: 100, 200, 400,
800, 1200, 1600 and 2000~K. From the H$_2$O$-$H$_2$ data of Phillips
et al. (\cite{phillips96}), we observed the temperature dependence of
de-excitation rates below 100~K to be very weak: de-excitation rates
at 20~K are always within a factor of 2 of those at 100~K. This is
clearly illustrated in Fig.~1. Below 100~K, a flat temperature
dependence of the de-excitation rate was thus assumed, except for the
lowest five para- and five ortho-H$_2$O levels for which the values at
20~K were taken from the quantum CC results of Dubernet et
al. (\cite{dubernet06}). Such a flat low temperature dependence
actually reflects the influence of the deep potential well of the
H$_2$O$-$H$_2$ PES ($\sim$ 235~cm$^{-1}$) which supports large shape
resonances, in contrast to H$_2$O$-$He ($\sim$
35~cm$^{-1}$). Obviously, the above low temperature extrapolation
procedure introduces errors but these are believed not to exceed a
factor of 2$-$3 for the largest rates. The complete set of
de-excitation rates among all 45 para- and ortho-H$_2$O levels between
20 and 2000~K are provided as online material and will be made
available in the BASECOL database ({\tt
http://www.obspm.fr/basecol/}). Excitation rates can be obtained from
the detailed balance relation, Eq.~(5).

In order to assess quantitatively the impact of the new H$_2$O$-$H$_2$
collisional rates on water emission, detailed radiative transfer
studies adapted to various astronomical environments are required. In
Sect.~4 below, this impact is investigated for a range of density and
temperature conditions by neglecting all excitation mechanisms other
than collisions. A first-order indicator is also provided by the
impact of rates on critical densities. For a multi-level system, the
critical density of a given partner (here para- or ortho-H$_2$) is
usually defined, in the optically thin case, given that the density at
which the sum of the collisional de-excitation rates of a given level
is equal to the sum of the spontaneous radiative de-excitation rates:
\begin{equation}
n_{cr}(J_{\alpha},T)=\frac{\sum_{J'_{\alpha'}}A(J_{\alpha}\rightarrow
J'_{\alpha'})} {\sum_{J'_{\alpha'}}R(J_{\alpha}\rightarrow J'_{\alpha'}, T)}.
\end{equation}
When the colliding partner has a density $n\gg n_{cr}$, collisions
maintain rotational levels in LTE at the kinetic temperature while for
densities $n\lesssim n_{cr}$, deviations from LTE including population
inversions are expected. Eq.~(8) was computed with the present
collisional rates for para-H$_2$ and ortho-H$_2$ and with the scaled
H$_2$O$-$He rates of Green et al. (\cite{green93}). Einstein
coefficients were taken from the {\it Jet Propulsion Laboratory}
catalogue. Results are presented in Table~1 for 6 representative water
levels, namely the upper level of the 557~GHz transition
($1_{10}-1_{01}$) and the upper levels of maser transitions known to
be collisionally pumped (Yates et al. \cite{yates97}): the 183~GHz
line ($3_{13}-2_{20}$), the 380~GHz line ($4_{14}-3_{21}$), the
325~GHz line ($5_{15}-4_{22}$), the 22~GHz line ($6_{16}-5_{23}$) and
the 321~GHz line ($10_{29}-9_{36}$). Note that maser transitions are
expected to be quite sensitive to relatively small differences in
collisional rates. As illustrated in Table~1, critical densities for
H$_2$ are decreased with respect to the scaled He values by a typical
factor of 2$-$5 at 100~K and by a factor of 1$-$3 at higher
temperatures. These differences are still modest (with the exception
of $10_{29}$) because collisional rates are summed over all possible
downward transitions, thus averaging the impact of the new
rates. Larger differences in the emission line fluxes are however
expected as individual state-to-state collisional rates, in particular
with ortho-H$_2$, can exceed the scaled He values by more than an
order of magnitude. This is demonstrated in Sect.~4. We conclude that
using the new H$_2$O$-$H$_2$ collisional rates, non-LTE effects
including population inversions will be quenched at lower H$_2$
densities, especially in regions where the H$_2$ ortho/para ratio is
large.

\begin{table}
\caption{Critical densities, in cm$^{-3}$, for (scaled) He, para-H$_2$
and ortho-H$_2$, as functions of temperature and for a set of
representative H$_2$O levels. Powers of 10 are given in
parentheses. See text for details.}
\label{table:1}    
\centering                         
\begin{tabular}{l c c c c c c}       
\hline\hline
$J_{K_aK_c}$ & $T$(K) & He & p-H$_2$ & o-H$_2$ & He/p-H$_2$ & He/o-H$_2$ \\
\hline
$1_{10}$  & 100  & 1.3(08)  & 6.3(07) & 3.0(07) & 2.1 & 4.3   \\      
          & 300  & 5.7(07)  & 3.6(07) & 2.5(07) & 1.6 & 2.3   \\      
          & 1000 & 3.0(07)  & 2.3(07) & 2.2(07) & 1.3 & 1.4   \\      
 & & & & & & \\
$3_{13}$  & 100  & 3.2(09)  & 1.1(09) & 7.0(08) & 2.9 & 4.6   \\      
          & 300  & 1.6(09)  & 7.7(08) & 5.6(08) & 2.1 & 2.9   \\      
          & 1000 & 8.4(08)  & 5.1(08) & 4.5(08) & 1.6 & 1.9   \\      
 & & & & & & \\
$4_{14}$  & 100  & 6.6(09)  & 2.5(09) & 1.4(09) & 2.6 & 4.7   \\      
          & 300  & 3.2(09)  & 1.5(09) & 1.1(09) & 2.1 & 2.9   \\      
          & 1000 & 1.5(09) & 9.3(08) & 8.2(08) & 1.6 & 1.8   \\      
 & & & & & & \\
$5_{15}$  & 100  & 1.2(10)  & 4.8(09) & 2.6(09) & 2.5 & 4.6   \\      
          & 300  & 5.8(09)  & 2.9(09) & 1.9(09) & 2.0 & 3.1   \\      
          & 1000 & 2.5(09)  & 1.6(09) & 1.4(09) & 1.6 & 1.8   \\      
 & & & & & & \\
$6_{16}$  & 100  & 2.3(10)  & 8.6(09) & 4.2(09) & 2.7 & 5.5   \\      
          & 300  & 1.0(10)  & 4.9(09) & 3.1(09) & 2.0 & 3.2   \\      
          & 1000 & 4.1(09)  & 2.5(09) & 2.1(09) & 1.6 & 2.0   \\ 
 & & & & & & \\
$10_{29}$ & 100  & 2.0(11)  & 4.0(10) & 9.8(09) & 5.0 & 20.4 \\      
          & 300  & 6.9(10)  & 1.9(10) & 8.1(09) & 3.6 & 8.5  \\      
          & 1000 & 2.0(10)  & 8.5(09) & 6.3(09) & 2.4 & 3.2  \\     
\hline                  
\end{tabular}
\end{table}

\section{Modelling of water emission}

In this section we explore, by means of a LVG code, the impact of the
new computed collisional coefficients on the theoretical predictions
of water line emission.  The code is adapted from the one described in
Ceccarelli et al. (\cite{ceccarelli02}). It refers to a semi-infinite
isodense and isothermal slab (plane-parallel geometry). The code
solves self-consistently the statistical level populations and the
radiative transfer equations under the approximation of the escape
probability, assuming a gradient in the velocity field (Capriotti
\cite{capriotti65}). We consider the first 45 levels of both ortho-
and para-H$_2$O separately. The computed line spectrum depends on a
few basic parameters: the density of the colliders (H$_2$), the
temperature of the emitting gas, the water column density and the
maximum gas velocity.  To have a relatively exhaustive study, we
explored a large parameter space. We varied the H$_2$ density between
10$^{4}$ and 10$^{10}$ cm$^{-3}$; the kinetic temperature between 20
and 2000~K; the water column density between 10$^{10}$ and
10$^{15}$cm$^{-2}$, keeping the velocity equal to 1~km/s. Note that
the velocity only enters in the line opacity, coupled with the H$_2$O
column density. In the present study we considered specifically the
case of optically thin lines (the low H$_2$O column density case), and
the case where the lines are optically thick (the high H$_2$O column
density case). In both cases, depending on the gas density, the
collisional coefficients may play a major role in the predicted
flux. Indeed, even in the case of strongly optically thick lines, the
lines could be ``effectively optically thin'' if the levels are very
sub-thermally populated, as is often the case for water
lines. Finally, since the scope of this study is to understand the
impact of the newly computed collisional coefficients in the line
predictions, compared to the old ones, we did not consider the pumping
of the levels by infrared and submillimetre radiation, which would
unnecessarily complicate the problem.

\begin{figure*}[h]
  \centering
  \resizebox{\hsize}{!}{\includegraphics*[angle=-90]{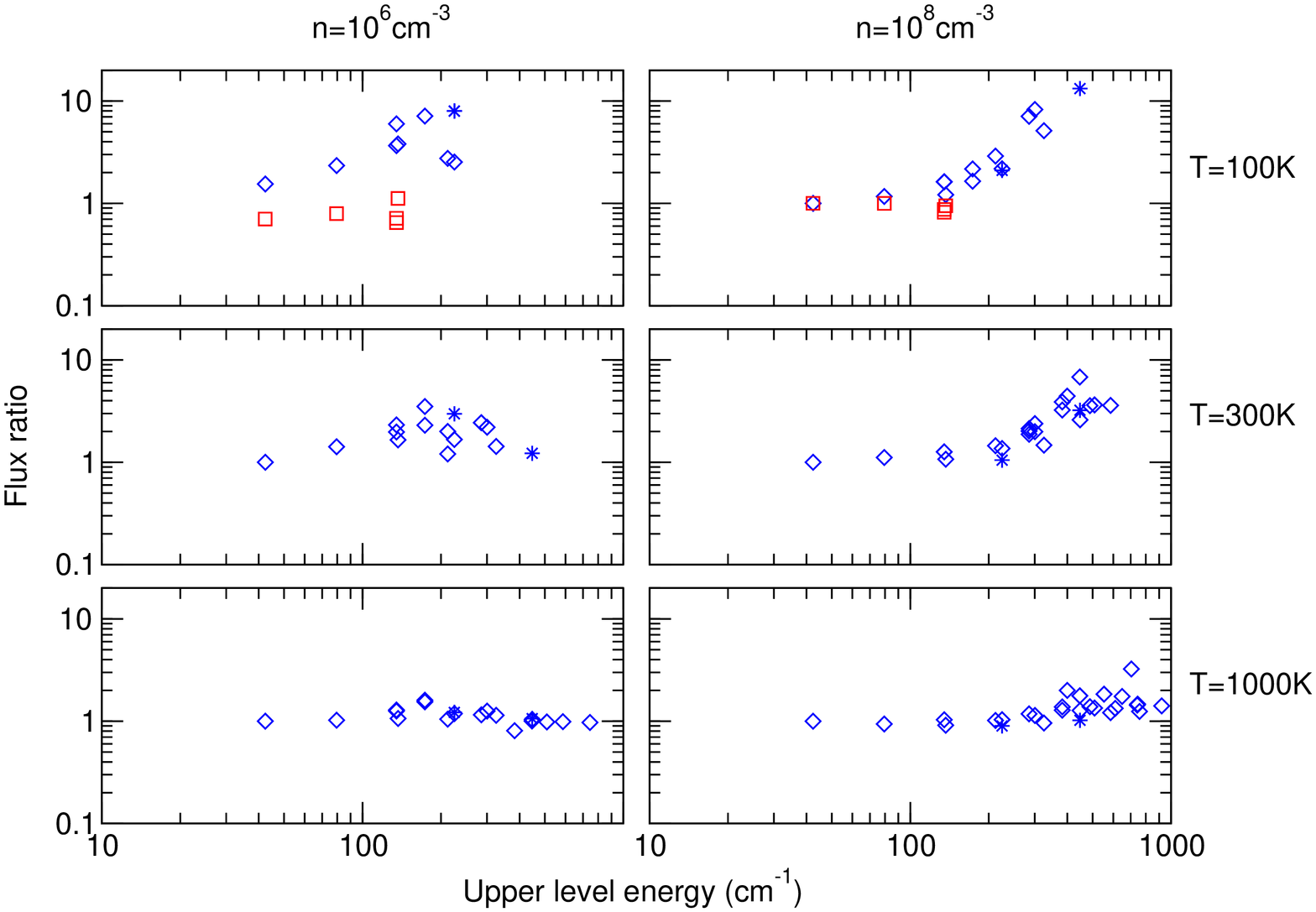}}
  \caption{Emission line flux ratios from LVG modelling as functions
  of the upper energy levels of ortho-H$_2$O for representative
  densities and temperatures. Three differents sets of collision rates
  were employed (see text for details). The H$_2$O column density is
  fixed at 10$^{15}$~cm$^{-2}$. Squares denote the ratios between LVG
  fluxes based on the classical rates and LVG fluxes based on the
  H$_2$O$-$H$_2$ rates of Phillips et
  al. (\cite{phillips96}). Diamonds and stars give the ratio between
  LVG fluxes based on the classical rates and LVG fluxes based on the
  scaled H$_2$O$-$He rate of Green et al. (\cite{green93}). Diamonds
  correspond to transitions whose fluxes are larger than 1\% of the
  total flux; stars correspond to the radio transitions at 22 and
  380~GHz.}
  \label{Figrates2}
\end{figure*}

We compared the results of the LVG computations with three sets of
collisional data: the scaled H$_2$O-He rates of Green et
al. (\cite{green93}) between 20 and 2000~K, the H$_2$O$-$H$_2$ rates
of Phillips et al. (\cite{phillips96}) between 20 and 140~K and the
present H$_2$O$-$H$_2$ classical rates beween 20 and 2000~K. The
results of this comparison are reported in Fig. \ref{Figrates2}, where
we show the line flux ratios for ortho-H$_2$O at representative
densities and temperatures. Collision rates with ortho-H$_2$ only are
considered to simplify the interpretation. The H$_2$O column density
is fixed at 10$^{15}$~cm$^{-2}$. It can be observed that LVG fluxes
based on the classical rates are increased up to a factor of 10 with
respect to LVG fluxes based on the scaled H$_2$O$-$He rates of Green
et al. (\cite{green93}). Even larger differences were observed at
lower temperatures. Second, LVG fluxes based on the classical rates
are very similar to those based on the rates of Phillips et
al. (\cite{phillips96}), with flux ratios close to 1. These findings
clearly reflect the differences in collision rates, such as those
illustrated in Fig.~1. They also show that differences in rates are
not amplified within the radiative transfer equations. In particular,
it is worth noting that the typical increase of fluxes at 1000~K is
less than a factor of 2 at the investigated densities. Similar results
were obtained with para-H$_2$, especially above 300~K where
differences between para- and ortho-H$_2$ rates are minor. We note,
however, that below 100~K, the classical para-H$_2$ rates are not
necessarily more accurate than the scaled H$_2$O$-$He rates. Third, we
observe that the flux ratios rise steeply with increasing upper
energies. This is expected since the critical densities are larger for
higher levels and, therefore, the impact of collision rates,
i.e. non-LTE effects, becomes more pronounced. Finally, when the
temperature increases, the impact of collision rates is reduced,
partly because the low lying levels become thermalized and partly
because the differences in rates decrease. It should be noted,
however, that at densities higher than 10$^{8}$cm$^{-3}$, large flux
ratios ($\sim$ 10) were still observed at 1000~K for high lying
levels.

We also tested the influence of the scaled H$_2$O$-$He rates employed
as substitutes for H$_2$ rates in the case of classically-forbidden
transitions. These rates, which generally correspond to $\Delta J>3$
and are lower than 10$^{-12}$ cm$^{3}$s$^{-1}$, were found to play a
minor role, especially at low temperatures ($T\leq$300~K). This
suggests that the radiative transfer equations are above all sensitive
to the largest rates. These findings however merit further
investigation.

\section{Conclusions}

We have computed classical rate coefficients for rotational
de-excitation among the lowest 45 para and 45 ortho rotational levels
of H$_2$O colliding with both para and ortho-H$_2$ in the temperature
range 20$-$2000~K. Trajectories were run on a recent high-accuracy
H$_2$O$-$H$_2$ PES using a canonical Monte-Carlo sampling. H$_2$
molecules were assumed to be rotationally thermalized at the kinetic
temperature. By comparison with quantum calculations available for low
lying levels, classical rates were found to be accurate within a
factor of 1$-$3 for the dominant transitions, i.e. those with rates
larger than a few 10$^{-12}$cm$^{3}$s$^{-1}$. Large differences with
the scaled H$_2$O$-$He rates of Green et al. (\cite{green93}) were
observed and emission line fluxes were shown to be increased by up to
a factor of 10. The present classical rates should therefore be
adopted in any detailed population model of water in warm and hot
environments.

It should be noted, however, that quantum CC calculations are
currently in progress (Dubernet and co-workers) and quantum rates will
ultimately replace the classical rates. It will thus be possible in
future studies to further test the sensitivity of water emission to
uncertainties in the collision rates and to identify, for example,
physical regimes where classical rates are of sufficient accuracy. We
also stress that the QCT method is currently the only practical
approach to investigate very high rotational angular momenta. For
example, water in supergiant stars has been observed through levels
with rotational quantum number $J$ up to 23 (e.g. Jennings \& Sada
\cite{jennings98}). In order to assess non-LTE effects in such
environments, classical calculations might provide precious
information. Finally, we note that the first excited vibrational level
of water lies only 2300~K above the ground state and it has been
detected in several sources (e.g. Menten et al. \cite{menten06}). Rate
coefficients for the total vibrational relaxation of H$_2$O by H$_2$
molecules have been computed by Faure et al. (\cite{faure05a}) and
some ``$J$-resolved'' cross sections can be found in Faure et
al. (\cite{faure05b}). Rovibrational state-to-state rates are however
unknown and will be considered in future works.

\begin{acknowledgements}
      All QCT calculations were performed on the {\it Service Commun
 de Calcul Intensif de l'Observatoire de Grenoble} with the valuable
 help from F. Roch. This research was supported by the CNRS national
 program "Physique et Chimie du Milieu Interstellaire" and by the FP6
 Research Training Network "Molecular Universe" (contract number
 MRTN-CT-2004-512302).

\end{acknowledgements}


\begin{thebibliography}{}

\bibitem[2002]{bala02} Balakrishnan, N., Yan, M., \& Dalgarno, A.\
2002, \apj, 568, 443

\bibitem[2002]{bergin02} Bergin, E.~A., \& Snell, R.~L.\ 2002, \apjl,
581, L105

\bibitem[1965]{capriotti65} Capriotti, E.~R.\ 1965, \apj, 142, 1101

\bibitem[1996]{ceccarelli96} Ceccarelli, C., Hollenbach, D.~J., \&
Tielens, A.~G.~G.~M.\ 1996, \apj, 471, 400

\bibitem[2002]{ceccarelli02} Ceccarelli, C., et al.\ 2002, \aap, 383,
603

\bibitem[2002]{cecchi02} Cecchi-Pestellini, C., Bodo, E.,
Balakrishnan, N., \& Dalgarno, A.\ 2002, \apj, 571, 1015

\bibitem[1990]{cernicharo90} Cernicharo, J., Thum, C., Hein, H., John,
D., Garcia, P., \& Mattioco, F.\ 1990, \aap, 231, L15

\bibitem[1999]{cernicharo99} Cernicharo, J., Pardo, J.~R.,
Gonz{\'a}lez-Alfonso, E., Serabyn, E., Phillips, T.~G., Benford,
D.~J., \& Mehringer, D.\ 1999, \apjl, 520, L131

\bibitem[2005]{cernicharo05} Cernicharo, J., \& Crovisier, J.\ 2005,
Space Science Reviews, 119, 29

\bibitem[2006]{cernicharo06} Cernicharo, J., Pardo, J.~R., \& Weiss,
A.\ 2006, \apjl, 646, L49

\bibitem[1969]{cheung69} Cheung, A.~C., Rank, D.~M., Townes, C.~H.,
Thornton, D.~D., \& Welch, W.~J.\ 1969, \nat, 221, 626


\bibitem[1997]{doty97} Doty, S.~D., \& Neufeld, D.~A.\ 1997, \apj,
489, 122

\bibitem[2002]{dubernet02} Dubernet, M.-L., \& Grosjean, A.\ 2002,
\aap, 390, 793

\bibitem[2006]{dubernet06} Dubernet, M.-L., Daniel, F., Grosjean A.,
et al.\ 2006, \aap, 460, 323

\bibitem[1993]{dutta93} Dutta, J.~M., Jones, C.~R., Goyette, T.~M., \&
de Lucia, F.~C.\ 1993, Icarus, 102, 232

\bibitem[2004]{faure04} Faure, A., Gorfinkiel, J.~D., \& Tennyson, J.\
2004, \mnras, 347, 323

\bibitem[2004]{faure04b} Faure, A., \& Wiesenfeld, L.\ 2004, \jcp,
121, 6771

\bibitem[2005a]{faure05a} Faure, A., Valiron, P., Wernli, M.,
Wiesenfeld, L., Rist, C., Noga, J., \& Tennyson, J.\ 2005a, \jcp, 122,
1102

\bibitem[2005b]{faure05b} Faure, A., Wiesenfeld, L., Wernli, M., \&
Valiron, P.\ 2005b, \jcp, 123, 4309

\bibitem[2006]{faure06} Faure, A., Wiesenfeld, L., Wernli, M., \&
Valiron, P.\ 2006, \jcp, 124, 214310

\bibitem[1998]{flower98} Flower, D.~R., Roueff, E., \& Zeippen, C.~J.\
1998, J. Phys. B: At. Mol. Opt. Phys., 31, 1105

\bibitem[2006]{gilijamse06} Gilijamse, J.~J., Hoekstra, S., van de
Meerakker, S.~Y.~T., Groenenboom, G.~C., Meijer, G., Science, 313, 1617

\bibitem[2007]{goddi07} Goddi, C., Moscadelli, L., Sanna, A.,
Cesaroni, R., \& Minier, V.\ 2007, \aap, 461, 1027

\bibitem[1993]{green93} Green, S., Maluendes, S., \& McLean, A.~D.\
1993, \apjs, 85, 181

\bibitem[2003]{grosjean03} Grosjean, A., Dubernet, M.-L., \&
Ceccarelli, C.\ 2003, \aap, 408, 1197


\bibitem[2003]{hjalmarson03} Hjalmarson, {\AA}., Frisk, U., Olberg,
M., et al.\ 2003, \aap, 402, L39

\bibitem[1998]{jennings98} Jennings, D.~E., \& Sada, P.~V.\ 1998,
Science, 279, 844

\bibitem[2006]{kondratko06} Kondratko, P.~T., Greenhill, L.~J., \&
Moran, J.~M.\ 2006, \apj, 652, 136

\bibitem[2003]{lefloch03} Lefloch, B., Cernicharo, J., Cabrit, S.,
Noriega-Crespo, A., Moro-Mart{\'{\i}}n, A., \& Cesarsky, D.\ 2003,
\apjl, 590, L41

\bibitem[2004]{mandy04} Mandy, M.~E., \& Pogrebnya, S.~K.\ 2004, \jcp,
120, 5585

\bibitem[2005]{melnick05} Melnick, G.~J., \& Bergin, E.~A.\ 2005,
Advances in Space Research, 36, 1027

\bibitem[2006]{menten06} Menten, K.~M., Philipp, S.~D., G{\"u}sten,
R., Alcolea, J., Polehampton, E.~T., \& Br{\"u}nken, S.\ 2006, \aap,
454, L107

\bibitem[1991]{neufeld91} Neufeld, D.~A., \& Melnick, G.~J.\ 1991,
\apj, 368, 215

\bibitem[2006]{neufeld06} Neufeld, D.~A., Melnick, G~J.,
Sonnentrucker, P., et al.\ 2006, \apj, 649, 816

\bibitem[1994]{phillips94} Phillips, T.~R., Maluendes, S., McLean,
A.~D., \& Green, S.\ 1994, \jcp, 101, 5824

\bibitem[1996]{phillips96} Phillips, T.~R., Maluendes, S., \& Green,
S.\ 1996, \apjs, 107, 467

\bibitem[2005]{poelman05} Poelman, D.~R., \& Spaans, M.\ 2005, \aap,
440, 559

\bibitem[2007]{poelman07} Poelman, D.~R., Spaans, M., \& Tielens,
A.~G.~G.~M.\ 2007, \aap, 464, 1023

\bibitem[2007]{wernli07} Wernli, M., Wiesenfeld, L., Faure, A., \&
Valiron, P.\ 2007, \aap, 464, 1147

\bibitem[1998]{wolniewicz98} Wolniewicz, L., Simbotin, I., \&
Dalgarno, A.\ 1998, \apjs, 115, 293

\bibitem[1997]{yates97} Yates, J.~A., Field, D., \& Gray, M.~D.\ 1997,
\mnras, 285, 303

\end{thebibliography}
\end{document}